\documentclass[a4paper,12pt]{article}
\pdfoutput=1 

\usepackage{jcappub} 

\usepackage{bm}
\usepackage{indentfirst}
\usepackage{amsmath}
\usepackage{graphicx}
\usepackage{float}
\usepackage{amssymb}
\usepackage{subfigure}
\usepackage{amssymb}
\usepackage{psfrag}
\usepackage{epstopdf}
\usepackage{diagbox}

\title{NANOGrav signal and LIGO-Virgo Primordial Black Holes from the Higgs field}

\author{Zhu Yi$^{1, 2}$}
\author{and Zong-Hong Zhu$^{1,2}$}

\affiliation{$^1$Advanced Institute of Natural Sciences, Beijing Normal University, Zhuhai 519087, China}

\affiliation{$^2$Department of Astronomy, Beijing Normal University, Beijing 100875, China}

\emailAdd{yz@bnu.edu.cn}
\emailAdd{zhuzh@bnu.edu.cn}

\abstract{We show that the NANOGrav signal can come from the  Higgs field with a  noncanonical kinetic term
	in terms of the scalar induced gravitational waves.
	The scalar induced gravitational waves generated in our model are
	also detectable by space-based gravitational wave observatories.
	Primordial black holes with stellar masses that can explain LIGO-Virgo events are also produced.
	Therefore, the NANOGrav signal and the BHs in LIGO-Virgo events  may both originate from the Higgs field.}

\begin{document}
	\maketitle
	\flushbottom
	
	\section{Introduction}
	The North American Nanohertz Observatory for Gravitational Wave (NANOGrav)  Collaboration
	has recently published an analysis of the $12.5$ yrs pulsar timing array (PTA) data,
	where  strong evidence of a stochastic process  with a common amplitude and
	a common spectral slope across  pulsars was found \cite{NANOGrav:2020bcs}.
	Although this  process lacks quadrupolar spatial correlations,
	which should exist for gravitational wave (GW) signals,
	it is worthwhile to be interpreted as a  stochastic GW signal.
	The GW signal with the amplitude of the energy density $\Omega_\text{GW}h^2\sim 10^{-10}$
	at a reference frequency of $f_\text{yr}=1 \text{yr}^{-1}$ Hz
	has a flat power spectrum $\Omega_\text{GW}h^2\sim f^{\alpha}$
	with  $\alpha$ from $-1.5$ to $0.5$  at $1\sigma$ confidence level.

	The scalar induced gravitational waves (SIGWs) associated with the formation of primordial black holes (PBHs) are
	the natural sources to explain the NANOGrav signal
	\cite{DeLuca:2020agl,Inomata:2020xad, Vaskonen:2020lbd,Kohri:2020qqd,Domenech:2020ers,
		Vagnozzi:2020gtf,Kawasaki:2021ycf}.
	(For other explanations for the  sources of
	the NANOGrav signal, see Ref. \cite{Blasi:2020mfx,Ellis:2020ena,Li:2021qer,Atal:2020yic,Chen:2021wdo,Ramberg:2020oct,
		Middleton:2020asl,Li:2020cjj,Cai:2020qpu,Bhattacharya2021,
		Pandey:2020gjy,Bian:2021lmz,Sharma:2021rot,Kuroyanagi2021,Ratzinger:2020koh,Samanta:2020cdk}.)
	The nanohertz frequencies of SIGWs constrain the masses of PBHs to stellar mass,
	and these  stellar mass PBHs, on the other hand, may be the sources of the GWs detected
	by the Laser  Interferometer Gravitational Wave Observatory (LIGO)
	Scientific  Collaboration and the Virgo Collaboration
	\cite{Bird:2016dcv,Sasaki:2016jop, Abbott:2016blz,Abbott:2016nmj, Abbott:2017vtc,
		Abbott:2017oio,TheLIGOScientific:2017qsa,
		Abbott:2017gyy,LIGOScientific:2018mvr,Abbott:2020uma,LIGOScientific:2020stg,
		Abbott:2020khf,Abbott:2020tfl,LIGOScientific:2020ibl}.
	Therefore, PBHs with  stellar mass can  be related to both the LIGO-Virgo events and the NANOGrav signal.
	The PBHs  are also proposed to account for the dark matter (DM)
	\cite{Ivanov:1994pa,Frampton:2010sw,Belotsky:2014kca,Khlopov:2004sc,Clesse:2015wea,
		Carr:2016drx,Inomata:2017okj,Garcia-Bellido:2017fdg,Kovetz:2017rvv,Pi:2017gih,Cai:2018dig,Carr:2020xqk}
	and explain  the Planet 9 which is a hypothetical astrophysical object
	in the outer solar system used to interpret
	the anomalous orbits of trans-Neptunian objects  \cite{Scholtz:2019csj}.
	
	PBHs will be formed from the  gravitational collapse if the density contrasts of overdense regions
	exceed the threshold value at the horizon reentry during radiation domination \cite{Carr:1974nx,Hawking:1971ei}.
	The initial conditions for the overdense regions are from the  inflation.
	Enough abundance of PBHs needs the  amplitude of the power spectrum of
	the primordial curvature perturbations  reach $A_\zeta\sim \mathcal{O}(0.01)$ at small scales,
	and this condition is also required to explain the NANOGrav signal if it is regarded as a SIGW.
	The constraint on the amplitude of the power spectrum at large scales from the
	cosmic microwave background (CMB) anisotropy measurements is $A_\zeta=2.1\times10^{-9}$ \cite{Planck:2018jri}.
	Therefore, to produce enough abundance of  PBH DM, the amplitude of the power spectrum
	should be enhanced at least seven orders of magnitude
	to reach the threshold at small scales \cite{Gong:2017qlj,Lu:2019sti,Sato-Polito:2019hws}.
	
	The traditional slow-roll inflation  is unable to enhance the power spectrum to produce PBHs
	while keeping the model consistent with the CMB constraints.
	To overcome this,  the ultra-slow-roll inflation
	\cite{Martin:2012pe,Motohashi:2014ppa,Yi:2017mxs} is then
	considered \cite{Garcia-Bellido:2017mdw,Germani:2017bcs,Motohashi:2017kbs,
		Ezquiaga:2017fvi,Gong:2017qlj,Ballesteros:2018wlw,Dalianis:2018frf,
		Kamenshchik:2018sig,Fu:2019ttf,Fu:2019vqc,Dalianis:2019vit,
		Lin:2020goi,Braglia:2020eai,Gundhi:2020zvb,Cheong:2019vzl,Gao:2021vxb,Lin:2021vwc,Zhang:2021rqs}.
	Among them, the inflation model with a noncanonical kinetic term can successfully enhance
	the power spectra,   produce PBHs, and generate
	SIGWs \cite{Yi:2020cut,Yi:2020kmq,Gao:2020tsa}.
	In this mechanism, there are no restrictions on the potential,
	both sharp and broad peaks in the power spectrum can be generate,
	the masses of the PBHs and the frequencies of the SIGWs can be adjusted as we want.
	Although any  potential can be applied in this mechanism,
	the most natural potential to drive the inflation may be the Higgs field,
	as it is the only scalar field in the standard model of particle physics
	and has been detected by the Large Hadron Collider \cite{ATLAS:2012yve,Chatrchyan:2012ufa,ParticleDataGroup:2018ovx}.
	In this paper, we will show that under this mechanism,
	the NANOGrav signal and the BHs in the  LIGO-Virgo events
	can both come from the Higgs field.
	
	The paper is organized as follows.
	In Sec. 2, we give a brief review of the PBHs and  SIGWs.
	We introduce our model and  produce  PBHs with stellar mass and generate
	SIGWs consistent with the NANOGrav signal in Sec. 3,
	We conclude the paper in Sec. 4.
	\section{The PBHs  and scalar induced GWs}
	If the energy density contrast of overdensity regions is  large enough during the radiation domination,
	the PBHs will be formed from gravitational collapse,
	and the seed of the  overdensity regions is from
	the primordial curvature perturbations generated during inflation.
	The mass fraction of the Universe that collapses to form PBHs at formation is
	\begin{equation}\label{beta}
		\beta=\frac{\rho_{\text{PBH}}}{\rho_b},
	\end{equation}
	where $\rho_b$ is the energy density of the background
	and $\rho_{\text{PBH}}$ is the  energy density of the PBHs at  formation, which can be obtained by the
	peak theory\cite{Bardeen:1985tr,Green:2004wb,Young:2014ana,Germani:2018jgr,Young:2020xmk,Gow:2020bzo},
	\begin{equation}
		\rho_{\text{PBH}}=\int_{\nu_c}^{\infty}M_{\text{PBH}}\mathcal{N}_{pk}(\nu)d\nu,
	\end{equation}
	where the number density of the PBHs at formation  is
	\begin{equation}\label{num:den}
		\mathcal{N}_{pk}(\nu)=\frac{1}{a^3}\frac{1}{(2\pi)^2}\left(\frac{\sigma_1}{\sqrt{3}\sigma_0}\right)^3
		\nu^3\exp\left(-\frac{\nu^2}{2}\right).
	\end{equation}
	$\nu_c=\delta_c/\sigma_0$ and $\delta_c$  is the threshold for the formation of PBHs,
	$\sigma_0$ is the variance of the smoothed density contrast and
	$\sigma_1$ is the moment of the smoothed density power spectrum with the definition,
	\begin{equation}\label{variance1}
		\sigma^2_n=\int_{0}^{\infty}\frac{dk}{k}k^{2n} T^2(k,R_H)W^2(k,R_H)\mathcal{P}_\delta(k).
	\end{equation}
	The relation between the power spectrum of the density contrast $\mathcal{P}_\delta$ and
	the  power spectrum of  primordial curvature  perturbations $\mathcal{P}_\zeta$ is
	\begin{equation}\label{rel:pp}
		\mathcal{P}_\delta(k)=\frac{4(1+w)^2}{(5+3w)^2}\left(\frac{k}{aH}\right)^4 \mathcal{P}_{\zeta}(k),
	\end{equation}
	with the state equation $w=1/3$ during the radiation domination.
	The window function we choose in this paper is the real space top-hat window function
	\begin{equation}\label{window:fun}
		W(k,R_H)=3\left[\frac{\sin\left(kR_H\right)-\left(kR_H\right)
			\cos\left(kR_H\right)}{\left(kR_H\right)^3}\right],
	\end{equation}
	with the  smoothed scale    $R_H\sim 1/aH$.
	The threshold $\delta_c$ is dependent on the choice of the window function
	and the shape of the density perturbation \cite{Young:2020xmk,Musco:2018rwt,Germani:2018jgr}.
	For the real space top-hat window function, we choose the threshold
	as $\delta_c=0.51$ \cite{Young:2019osy,Musco:2018rwt}.
	During radiation domination with constant degrees of freedom,    the transfer function is
	\begin{equation}\label{transfer}
		T(k,R_H)=3\left[\frac{\sin\left(\frac{kR_H}{\sqrt{3}}\right)-\left(\frac{kR_H}{\sqrt{3}}\right)
			\cos\left(\frac{kR_H}{\sqrt{3}}\right)}{\left({kR_H}/{\sqrt{3}}\right)^3}\right].
	\end{equation}
	The masses of the PBHs  obey the critical scaling law \cite{Choptuik:1992jv,Evans:1994pj,Niemeyer:1997mt},
	\begin{equation}\label{pbh:mass}
		M_{\text{PBH}}=\kappa M_H(\delta-\delta_c)^{\gamma},
	\end{equation}
	with $\kappa=3.3$ for the real space top-hat window function and $\gamma=0.36$
	in the radiation domination \cite{Choptuik:1992jv,Evans:1994pj}.
	The horizon mass related to the horizon scale is
	\begin{equation}\label{mass:h}
		M_H\approx 13\left(\frac{g_*}{106.75}\right)^{-1/6}\left(\frac{k}{10^6 \text{Mpc}^{-1}}\right)^{-2}M_\odot,
	\end{equation}
	where $g_*$ is the number of relativistic degrees of freedom at the formation.
	The density parameter of the PBHs expressed by the $\beta$ is \cite{Byrnes:2018clq}
	\begin{equation}\label{beta:omega}
		\Omega_\text{PBH}=\int_{M_\text{min}}^{M_\text{max}} d \ln M_H \left(\frac{M_{eq}}{M_H}\right)^{1/2}\beta(M_H),
	\end{equation}
	where we use the relation $\rho_b\propto a^{-4}$ and $\rho_\text{PBH}\propto a^{-3}$ during radiation domination,
	and  $M_{eq}=2.8\times 10^{17}M_{\odot}$ is the horizon mass at the matter-radiation equality.
	Because of $\beta(M_H)\rightarrow 0$ at the condition $M_H\rightarrow 0$ or $M_H\rightarrow\infty$,
	for the sake of simplicity, we choose the interval of the integration as  $M_\text{min}=0$ and $M_\text{max}=\infty$.
	The fraction of the PBHs in the dark matter is
	\begin{equation}\label{fpbh:tot}
		f_{\text{PBH}}=\frac{\Omega_{\text{PBH}}}{\Omega_\text{DM}}=\int f(M_\text{PBH}) d\ln M_\text{PBH},
	\end{equation}
	where the definition of the PBHs mass function is
	\begin{equation}\label{mass:func}
		f(M_\text{PBH})=\frac{1}{\Omega_{\text{DM}}}\frac{d \Omega_{\text{PBH}}}{d \ln M_{\text{PBH}}}.
	\end{equation}
	Combining Eq. \eqref{beta:omega} and Eq. \eqref{mass:func} and using the
	relation \eqref{pbh:mass}, the mass function becomes \cite{Byrnes:2018clq}
	\begin{equation}\label{fpbh:beta}
		\begin{split}
			f(M_\text{PBH})&=\frac{1}{\Omega_{\text{DM}}} \int_{M_\text{min}}^{M_\text{max}}\frac{d M_H}{M_H}
			\frac{M_\text{PBH}}{\gamma M_H} \sqrt{\frac{M_{eq}}{M_H}}\\
			&\times\frac{1}{3\pi} \left(\frac{\sigma_1}{\sqrt{3}\sigma_0 aH}\right)^3\frac{1}{\sigma_0^4}
			\left(\mu^{1/\gamma}+\delta_c\right)^3\\
			&\times \mu^{1/\gamma} \exp\left[-\frac{\left(\mu^{1/\gamma}+\delta_c\right)^2}{2\sigma_0^2}\right],
		\end{split}
	\end{equation}
	where $\mu=M_{\text{PBH}}/(\kappa M_H)$ and we have used $d\delta/d\ln M_{\text{PBH}}=\mu^{1/\gamma}/\gamma$.

	Associating with  the formation of PBHs,
	the large scalar perturbations induce  the gravitational waves during radiation domination.
	These SIGWs belonging to the stochastic background  can account for
	the NANOGrav signal with  frequencies around $10^{-9}$ Hz \cite{Vaskonen:2020lbd,DeLuca:2020agl,Inomata:2020xad}
	and can also  be detected by the space-based GW detectors like
	LISA \cite{Danzmann:1997hm,Audley:2017drz}, Taiji \cite{Hu:2017mde} and TianQin  \cite{Luo:2015ght}
	with  frequencies around $10^{-3}$ Hz  in the future.
	In  the cosmological background and the  Newtonian gauge and  neglecting the anisotropic
	stress,  the perturbed metric is
	\begin{equation}
		\begin{split}
			d s^2=&-a^2(\eta)(1+2\Phi)d\eta^2 \\
			&+a^2(\eta)\left[(1-2\Phi)\delta_{ij}+\frac12h_{ij}\right]d x^i d x^j,
		\end{split}
	\end{equation}
	where $\eta$ is the conformal time, $\Phi$ is the Bardeen potential.
	In the  Fourier space, the  tensor perturbations $h_{ij}$ can be expressed as
	\begin{equation}
		\label{hijkeq1}
		h_{ij}(\bm{x},\eta)=\int\frac{  d^3k}{(2\pi)^{3/2}} e^{i\bm{k}\cdot\bm{x}}
		[h_{\bm{k}}(\eta)e_{ij}(\bm{k})+\tilde{h}_{\bm{k}}(\eta)\tilde{e}_{ij}(\bm{k})],
	\end{equation}
	where $e_{ij}(\bm{k})$ and $\tilde{e}_{ij}(\bm{k})$  are the plus and cross polarization tensors,
	\begin{gather}
		e_{ij}(\bm{k})=\frac{1}{\sqrt{2}}\left[e_i(\bm{k})e_j(\bm{k})-\tilde{e}_i(\bm{k})\tilde{e}_j(\bm{k})\right], \\
		\tilde{e}_{ij}(\bm{k})=\frac{1}{\sqrt{2}}\left[e_i(\bm{k})\tilde{e}_j(\bm{k})+\tilde{e}_i(\bm{k})e_j(\bm{k})\right],
	\end{gather}
	with $\bm e\cdot \tilde{\bm e}=\bm e \cdot \bm{k}= \tilde{\bm e}\cdot\bm{k}$.
	Focusing on  the source at second order from the linear scalar perturbations,
	the tensor perturbations in the Fourier space with either polarization  satisfy
	\cite{Ananda:2006af,Baumann:2007zm}
	\begin{equation}
		\label{eq:hk}
		h''_{\bm{k}}+2\mathcal{H}h'_{\bm{k}}+k^2h_{\bm{k}}=4S_{\bm{k}},
	\end{equation}
	where a prime denotes the derivative with respect to the conformal time,
	$h'_{\bm{k}}=dh_{\bm{k}}/d\eta$, and $\mathcal{H}=a'/a $ is the conformal Hubble parameter.
	The second order   source from the linear scalar perturbations is
	\begin{equation}
		\label{hksource}
		\begin{split}
			S_{\bm{k}}=&\int \frac{d^3\tilde{k}}{(2\pi)^{3/2}}e_{ij}(\bm{k})\tilde{k}^i\tilde{k}^j
			\left[2\Phi_{\tilde{\bm{k}}}\Phi_{\bm{k}-\tilde{\bm{k}}} \phantom{\frac{1}{2}}+ \right.\\
			&\left.\frac{1}{\mathcal{H}^2} \left(\Phi'_{\tilde{\bm{k}}}+\mathcal{H}\Phi_{\tilde{\bm{k}}}\right)
			\left(\Phi'_{\bm{k}-\tilde{\bm{k}}}+\mathcal{H}\Phi_{\bm{k}-\tilde{\bm{k}}}\right)\right],
		\end{split}
	\end{equation}
	where $\Phi_{\bm{k}}$ is Bardeen potential in  Fourier space,
	and  related to the primordial curvature perturbation $\zeta_{\bm{k}}$ by the transfer function,
	\begin{equation}
		\Phi_{\bm{k}}=\frac{3+3w}{5+3w}T(k\eta) \zeta_{\bm{k}}.
	\end{equation}
	The power spectrum $\mathcal{P}_h(k,\eta)$ for the SIGWs is
	\begin{equation}
		\label{eq:pwrh}
		\langle h_{\bm{k}}(\eta)h_{\tilde{\bm{k}}}(\eta)\rangle
		=\frac{2\pi^2}{k^3}\delta^{(3)}(\bm{k}+\tilde{\bm{k}})\mathcal{P}_h(k,\eta),
	\end{equation}
	which is found to be \cite{Baumann:2007zm,Ananda:2006af,Kohri:2018awv,Espinosa:2018eve}
	\begin{equation}\label{ph}
		\begin{split}
			\mathcal{P}_h(k,\eta)=&
			4\int_{0}^{\infty}dv\int_{|1-v|}^{1+v}du \left[\frac{4v^2-(1-u^2+v^2)^2}{4uv}\right]^2\\ &\times I_{RD}^2(u,v,x)\mathcal{P}_{\zeta}(kv)\mathcal{P}_{\zeta}(ku),
		\end{split}
	\end{equation}
	where $u=|\bm{k}-\tilde{\bm{k}}|/k$, $v=\tilde{k}/k$, $x=k\eta$  and the integral kernel $I_{\text{RD}}$  is
	\begin{equation}
		\label{irdeq1}
		\begin{split}
			I_{\text{RD}}(u, v, x)=&\int_1^x dy\, y \sin(x-y)\{3T(uy)T(vy)\\
			&+y[T(vy)u T'(uy)+v T'(vy) T(uy)]\\
			&+y^2 u v T'(uy) T'(vy)\}.
		\end{split}
	\end{equation}
	The   energy density of the SIGWs  is \cite{Espinosa:2018eve,Lu:2019sti}
	\begin{equation}
		\label{SIGWs:gwres1}
		\begin{split}
			\Omega_{\mathrm{GW}}(k,\eta)=&\frac{1}{6}\left(\frac{k}{aH}\right)^2\int_{0}^{\infty}dv\int_{|1-v|}^{1+v}du \\
			&\times\left[\frac{4v^2-(1-u^2+v^2)^2}{4uv}\right]^2\\
			&\times\overline{I_{\text{RD}}^{2}(u, v, x)} \mathcal{P}_{\zeta}(kv)\mathcal{P}_{\zeta}(ku),
		\end{split}
	\end{equation}
	where $\overline{I_{\text{RD}}^{2}}$ is the oscillation time average of the integral kernel.
	After  formation,  the SIGWs  behave  like radiation, so  the energy density of the SIGWs at present is
	\begin{equation}\label{d}
		\Omega_{\mathrm{GW}}(k,\eta_0)=c_g\Omega_{r,0}  \Omega_{\mathrm{GW}}(k,\eta),
	\end{equation}
	where $\Omega_{r,0}$ is the energy density of radiation at present, and \cite{Vaskonen:2020lbd,DeLuca:2020agl}
	\begin{equation}\label{gwcg}
		c_g=0.387\left(\frac{g_{*,s}^4g_*^{-3}}{106.75}\right)^{-1/3}.
	\end{equation}
	\section{The model and results}
	To obtain enough abundance of PBH DM and induce secondary GWs with large energy density, the amplitude of the power spectrum of
	the primordial curvature perturbations should be enhanced about seven order of magnitude at small scales.
	In this section, we present our model and show that
	the PBHs that  account for LIGO-Virgo events and
	the SIGWs which are consistent with  the NANOGrav signal
	may come from the Higgs field with the form
	\begin{equation}\label{higgssm}
		V(H_\text{SM})=-\mu^2 H_\text{SM}^\dag H_\text{SM}+\lambda(H_\text{SM}^\dag H_\text{SM})^2,
	\end{equation}
	where $H_\text{SM}$ is the SM Higgs boson, $\lambda$ is the self-coupling constant, and the vacuum expectation value $v_\text{EW}\simeq 246\text{Gev}$. For the unitary gauge $H_\text{SM}=(0,\phi+v_\text{EW})^T/\sqrt{2}$, in the inflation epoch $\phi\gg v_\text{EW}$,  the potential \eqref{higgssm} becomes 
	\begin{equation}\label{higgs:p}
		V(\phi)=\frac{\lambda}{4}\phi^4.
	\end{equation}
	The model that can enhance the power spectrum  is
	\begin{equation}\label{act1}
		S=\int d x^4 \sqrt{-g}\left[\frac{1}{2}R+X+G(\phi)X-V(\phi)\right],
	\end{equation}
	where $X=-g_{\mu\nu}\nabla^{\mu}\phi\nabla^{\nu}\phi/2$  and we take the convention $8\pi G=1$. 
	The  noncanonical coupling function $G(\phi)$, which may arise from scalar-tensor theory of gravity,
	G inflation \cite{Kobayashi:2010cm} or k inflation \cite{Garriga:1999vw,ArmendarizPicon:1999rj}, is expressed as follows:
	\begin{equation}
		G(\phi)=G_p(\phi)+f(\phi).
	\end{equation}
Function $f(\phi)$ is related to the potential and used to make the model consistent with the observational data.   Function $G_p(\phi)$  has a high peak used to enhance the power spectrum, in this paper it is \cite{Yi:2020cut}
	\begin{equation}\label{gfuncng}
		G_p(\phi)=\frac{d}{1+\left(|\phi-\phi_p|/{w}\right)^q},
	\end{equation}
where $d$ gives the amplitude of the peak, 
$\phi_p$ determines the position of the peak in the power spectrum,  
and  $q$  controls the shape of the peak.
	
The background equations are
	\begin{gather}
		\label{Eq:eom1}
		3H^2=\frac{1}{2}\dot{\phi}^2+V(\phi)+\frac{1}{2}\dot{\phi}^2G(\phi),\\
		\label{Eq:eom2}
		\dot{H}=-\frac{1}{2}[1+G(\phi)]\dot{\phi}^2,\\
		\label{Eq:eom3}
		\ddot{\phi}+3H\dot{\phi}+\frac{V_{\phi}+\dot{\phi}^2G_{\phi}/2}{1+G(\phi)}=0,
	\end{gather}
	where $G_\phi=dG(\phi)/d\phi$ and $V_\phi=dV/d\phi$. 	
The quadratic action for the curvature perturbation $\zeta$ is \cite{Garriga:1999vw,Kobayashi:2010cm}
	\begin{equation}\label{quadratic:action}
		S^{(2)}=\frac{1}{2}\int d\tau d^3x\tilde{z}^2
		\left[\mathcal{G}(\zeta')^2-\mathcal{F}(\vec{\nabla}\zeta)^2\right],
	\end{equation}
	where the prime represents derivative with respect to the conformal time $\tau$ and
	$\tilde{z}=a\dot{\phi}/H$, $\mathcal{G}=\mathcal{F}=1+G$.
	Since the sound speed for the scalar mode is $c_s^2=\mathcal{F}/\mathcal{G}=1$,
	so there is no problem with ghost and gradient instabilities in this mechanism \cite{Lin:2020goi}. The equation for 
	the curvature perturbation is
	\begin{equation}
		\label{zeta:k}
		\frac{d^2u_k}{d\tau^2}+\left(k^2-\frac{1}{z}\frac{d^2z}{d\tau^2}\right)u_k=0,
	\end{equation} 
	where  $u_k=z\zeta_k$ and $z=a\dot{\phi}(1+G)^{1/2}/H$. Solving   Eq. \eqref{zeta:k}, we can obtain the 
	power spectrum for the curvature perturbation. Under the slow-roll conditions, we get
	\begin{equation}\label{curvature:perturbation}
		\mathcal{P}_{\zeta}\approx \frac{V^3}{12\pi^2V_\phi^2}[1+G_p(\phi)+f(\phi)].
	\end{equation}
	With the help of the peak function $G_p(\phi)$, the scalar power spectrum can be enhanced easily.
	
In our model, away from the peak,  the peak function $G_p$ can be neglected and the function $f(\phi)$ dominates, we can use the slow-roll conditions to obtain  the scalar tilt and  tensor-to-scalar ratio,
	\begin{gather}
		\label{scalartilt}
		n_s-1\approx \frac{1}{1+f}\left(2\eta_V-6\epsilon_V -\frac{f_{\phi}}{1+f}\sqrt{2\epsilon_V}\right),\\
		\label{tts}
		r \simeq \frac{16\epsilon_V}{1+f},
	\end{gather}
	where the slow-roll parameters are
	\begin{equation}\label{slp}
		\epsilon_V=\frac{1}{2}\left(\frac{V_{\phi}}{V}\right)^2, \quad \eta_V=\frac{V_{\phi\phi}}{V}.
	\end{equation} 
To analyze the scalar tilt and tensor-to-scalar ratio, we can  take the transformation 
\begin{equation}\label{para:trans}
	d\Phi=\sqrt{1+f(\phi)}d\phi, \quad U(\Phi)=V[\phi(\Phi)],
\end{equation}
changing the scalar field $\phi$ to the new field $\Phi$ with the new potential $ U(\Phi)$. In the new field and potential, 
the scalar tilt \eqref{scalartilt}  and tensor-to-scalar ratio \eqref{tts} become  the standard form,
\begin{gather}\label{can:scalartilt}
	n_s-1\simeq 2\eta_U-6\epsilon_U,\\
	\label{can:tts}
	r\simeq 16\epsilon_U,
\end{gather}
where the new slow-roll parameters are
\begin{equation}
	\epsilon_U=\frac{1}{2}\left(\frac{U_{\Phi}}{U}\right)^2, \quad \eta_U=\frac{U_{\Phi\Phi}}{U}.
\end{equation} 
The $e$-folding numbers are
\begin{equation}
	N=\int_{\Phi_e}^{\Phi_*}\frac{U}{U_\Phi}d\Phi +\Delta N,
\end{equation}
where the first term is the $e$-folding numbers from the standard slow-roll inflation and the second term $\Delta N\approx 25$  \cite{Yi:2020cut} is  from the peak function.
Since the peak function $G_p(\phi)$ contributes to about  $25$ $e$-folding number, 
the  effective $e$-folding number of  the standard equations \eqref{can:scalartilt} and \eqref{can:tts} is reduced to  $N_\text{eff}\approx 35$.

For $N_\text{eff} \approx 35$, to satisfy  the observational data, a suitable parameterization for the scalar tilt is 
\begin{equation}\label{para:ns}
		n_s=1-\frac{7}{6N_\text{eff}}.
\end{equation}
By the method of potential reconstruction \cite{Lin:2015fqa}, the reconstructed potential from the parameteriztion \eqref{para:ns} is
\begin{equation}\label{para:phi}
		U(\Phi)=U_0\Phi^{1/3},
\end{equation}
which gives 
\begin{equation}\label{para:r}
		r=\frac{4}{3N_\text{eff}}.
\end{equation}
	Combining equations \eqref{para:ns} and  \eqref{para:r}, for the  effective $e$-folding number $N_\text{eff}\approx 35$, the  potential \eqref{para:phi} gives 
	\begin{equation}\label{the:nsr}
		n_s=0.967,\quad r=0.038,
	\end{equation}
	which are consistent with the CMB observational constraints at large scales \cite{Planck:2018jri}.  
	
	Combining the   transformation \eqref{para:trans} and potential  \eqref{para:phi},  we can   obtain 
	\begin{equation} \label{f:potential}
		f(\phi)=9\left(\frac{1}{U_0}\right)^6 V(\phi)^4V_\phi^2.
	\end{equation}
Substituting the Higgs potential  \eqref{higgs:p} into the relation \eqref{f:potential}, we have 
	\begin{equation}
		f(\phi)=f_0 \phi^{22},\quad f_0=144\left(\frac{\lambda}{4U_0}\right)^6,
	\end{equation}
where we take  $f_0=1$ for simplicity. At the lower energy scales $\phi\ll 1$,  the non-minimal coupling function $G(\phi)$ is neglected and the model  reduces to the standard case with the canonical kinetic term and Higgs potential  \eqref{higgs:p}.
For other potential $V(\phi)$, such as the E-model \cite{Yi:2020cut} and natural inflation \cite{Gao:2020tsa}, we can also  find the corresponding function $f(\phi)$ and give the same potential $U(\Phi)$. Therefore, the inflationary potential does not have any restriction in our mechanism.
Combining potential \eqref{para:phi} and the transformation \eqref{para:trans}, we can obtain the scalar field at the horizon cross, 
	\begin{equation}
		\phi_*=1.4.
	\end{equation}   
Guided by equation \eqref{curvature:perturbation},   we take $\lambda=1.25\times 10^{-9}$ to satisfy the observational constraint $A_\zeta=2.1\times 10^{-9}$  \cite{Planck:2018jri}. To produce SIGWs with the peak around the  nanohertz frequencies, we take $\phi_p=1.367$ and $w=3.15\times 10^{-10}$ in the peak function \eqref{gfuncng}.  For the index in the peak function \eqref{gfuncng}, we take $q=7/5$,  so that the SIGWs generated in our models can  be also  detected by the future space-based GW detectors. 
	
\begin{figure}[htp]
		\centering
		\includegraphics[width=0.8\columnwidth]{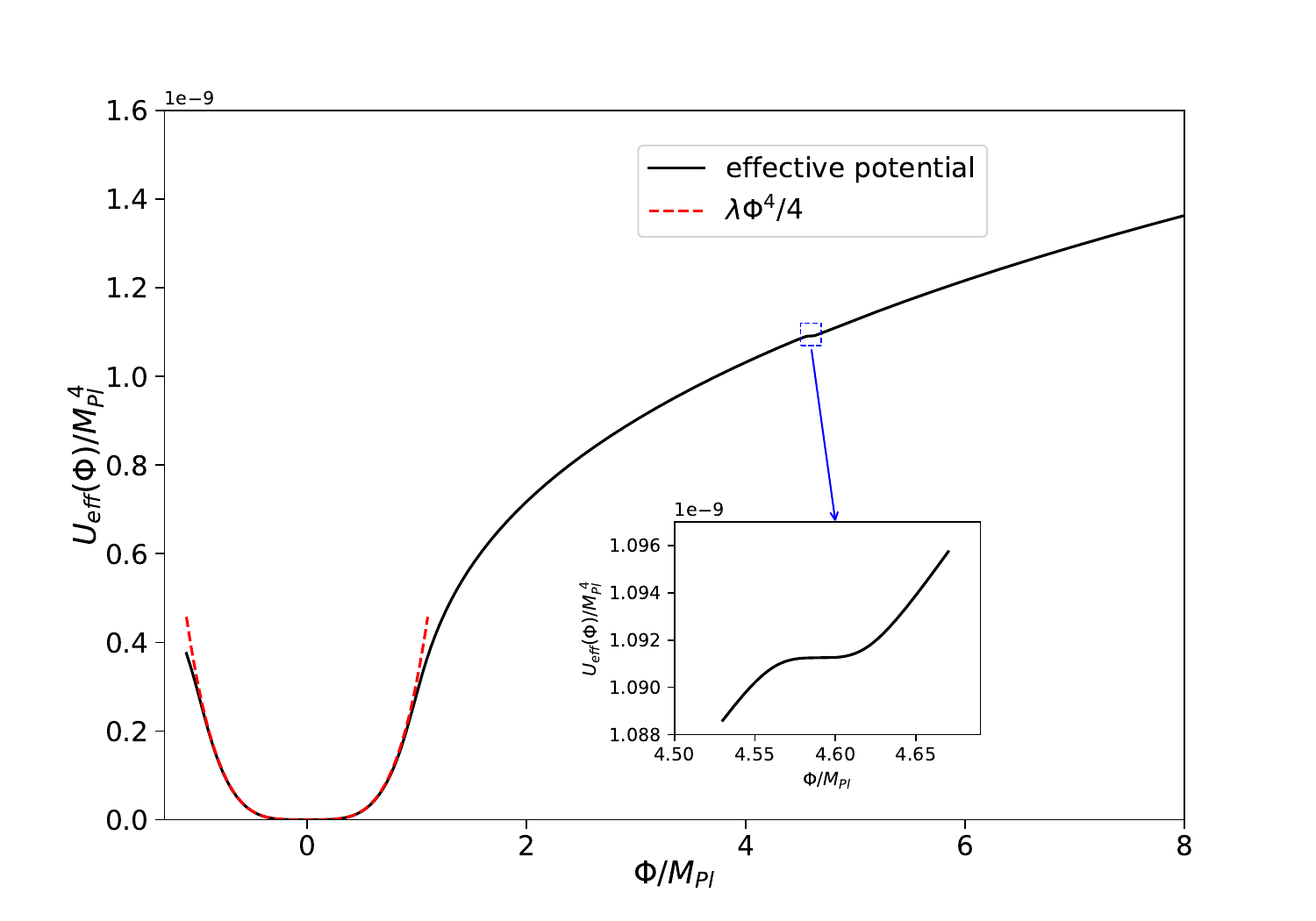}
	\caption{The  black solid  line denotes the effective potential $U_\text{eff}(\Phi)$ of the model with $d=3.23\times 10^{11}$,  and  $M_{Pl}=1/\sqrt{8\pi G}$ is the reduced Planck mass.  The red dashed  line denotes the potential $\lambda\Phi^4/4$.   At the low energy scales $\Phi/M_{Pl}\ll1$, the effective potential reduces to the Higgs potential.    }
		\label{potential:pbh}
	\end{figure}

We can also take the  transformation
	\begin{equation}
		d\Phi=\sqrt{1+G(\phi)}d\phi,
	\end{equation}
	to change the non-canonical field $\phi$ to be the canonical field $\Phi$ and get the effective potential $U_\text{eff}(\Phi)$ for the model \eqref{act1}. In general, it is hard to get the analytical relation between the non-canonical field $\phi$ and the canonical field $\Phi$ and to obtain the expression for  the effective potential $U_\text{eff}(\Phi)$.  However, we can use the numerical method to  obtain the effective potential. For the only undetermined parameter $d$ in the peak function \eqref{gfuncng}, we take 
	\begin{equation}\label{para:dq}
		d=3.23\times 10^{11},
	\end{equation}
as an example to numerically obtain the  effective potential which is shown in  Figure \ref{potential:pbh}. There is an inflection point in the effective potential $U_\text{eff}(\Phi)$ and the existence of the inflection point in the effective potential is the reason why our model can enhance the power spectrum.  
	
\subsection{The results of the PBHs}
In this section, we present the numerical results about the PBHs with the  parameter $d$ taking the value  in the equation \eqref{para:dq}.  
\begin{figure}[htbp]
		\centering
		\includegraphics[width=0.6\columnwidth]{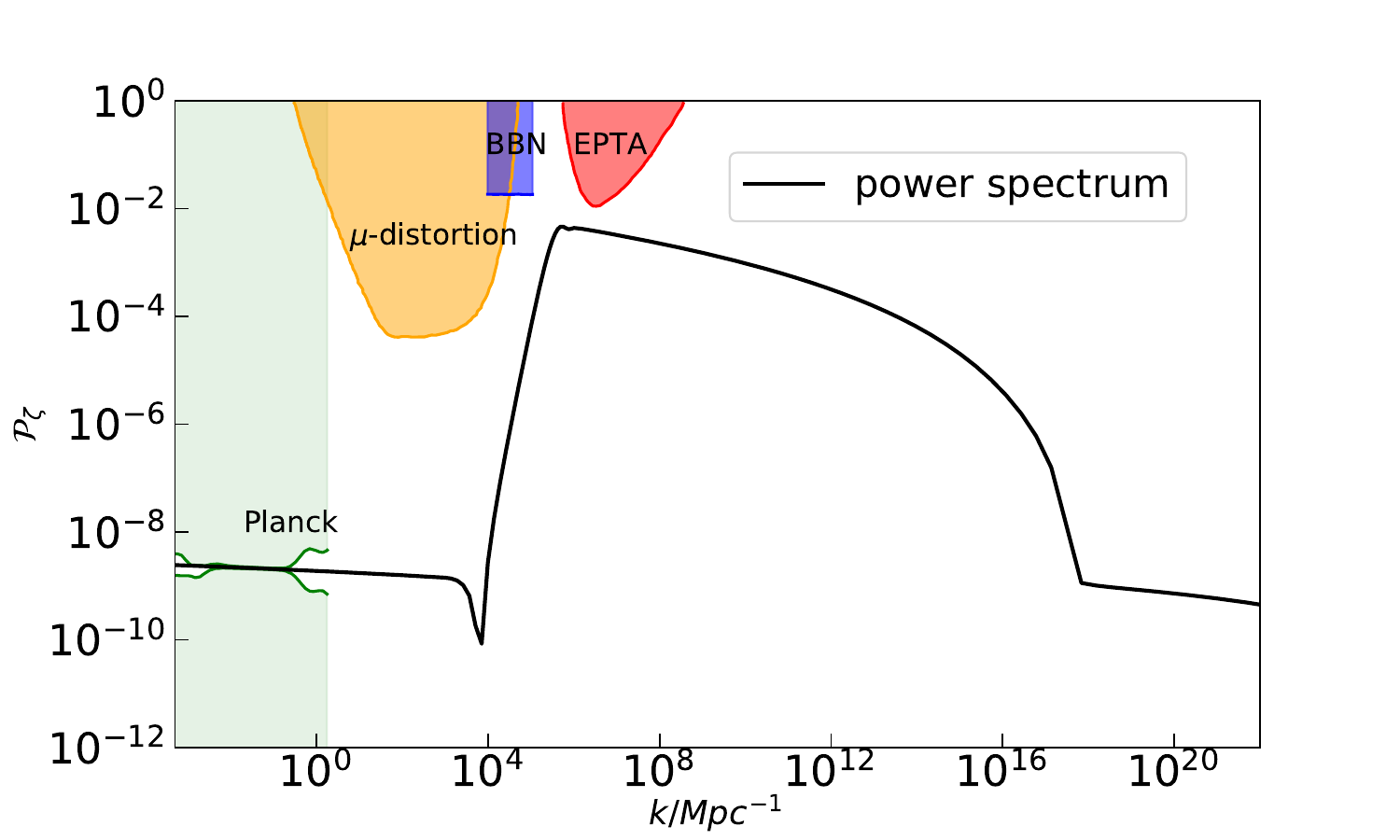}
		\caption{The  power spectrum from model \eqref{act1} for the case \eqref{para:dq}.
			The light green shaded region is excluded by the CMB observations \cite{Planck:2018jri}.
			The red, blue and orange regions show the
			constraints from the PTA observations \cite{Inomata:2018epa},
			the effect on the ratio between neutron and proton
			during the big bang nucleosynthesis (BBN) \cite{Inomata:2016uip}
			and $\mu$-distortion of CMB \cite{Fixsen:1996nj}, respectively.}
		\label{pic:pr}
	\end{figure}
	The numerical solutions for the scalar tilt and the tensor-to-scalar ratio are
	\begin{equation}\label{cmb:nsr}
		n_s=0.966,\quad r=0.039,
	\end{equation}
	which are compatible with the Planck 2018 constraints \cite{Planck:2018jri} and the theoretic result \eqref{the:nsr},
	and the $e$-folds before the end of
	inflation at the horizon exit for the pivotal scale
	$k_*=0.05 \text{Mpc}^{-1}$ are $N=63$. The numerical solution for the primordial  power spectrum is displayed in Figure \ref{pic:pr}, 
	which is  consistent with  the constraints from the PTA observations \cite{Inomata:2018epa},
	the effect on the ratio between neutron and proton during the big bang nucleosynthesis (BBN) \cite{Inomata:2016uip}
	and $\mu$-distortion of CMB \cite{Fixsen:1996nj}. The power spectrum at the peak and the peak scale are
	
	\begin{equation}\label{powerpeak}
		\mathcal{P}_{\zeta(\text{peak})}=4.63\times 10^{-3},\quad k_\text{peak}=5.65\times 10^5 \text{Mpc}^{-1}.
	\end{equation}

\begin{figure}[htp]
		\centering
		\includegraphics[width=0.65\columnwidth]{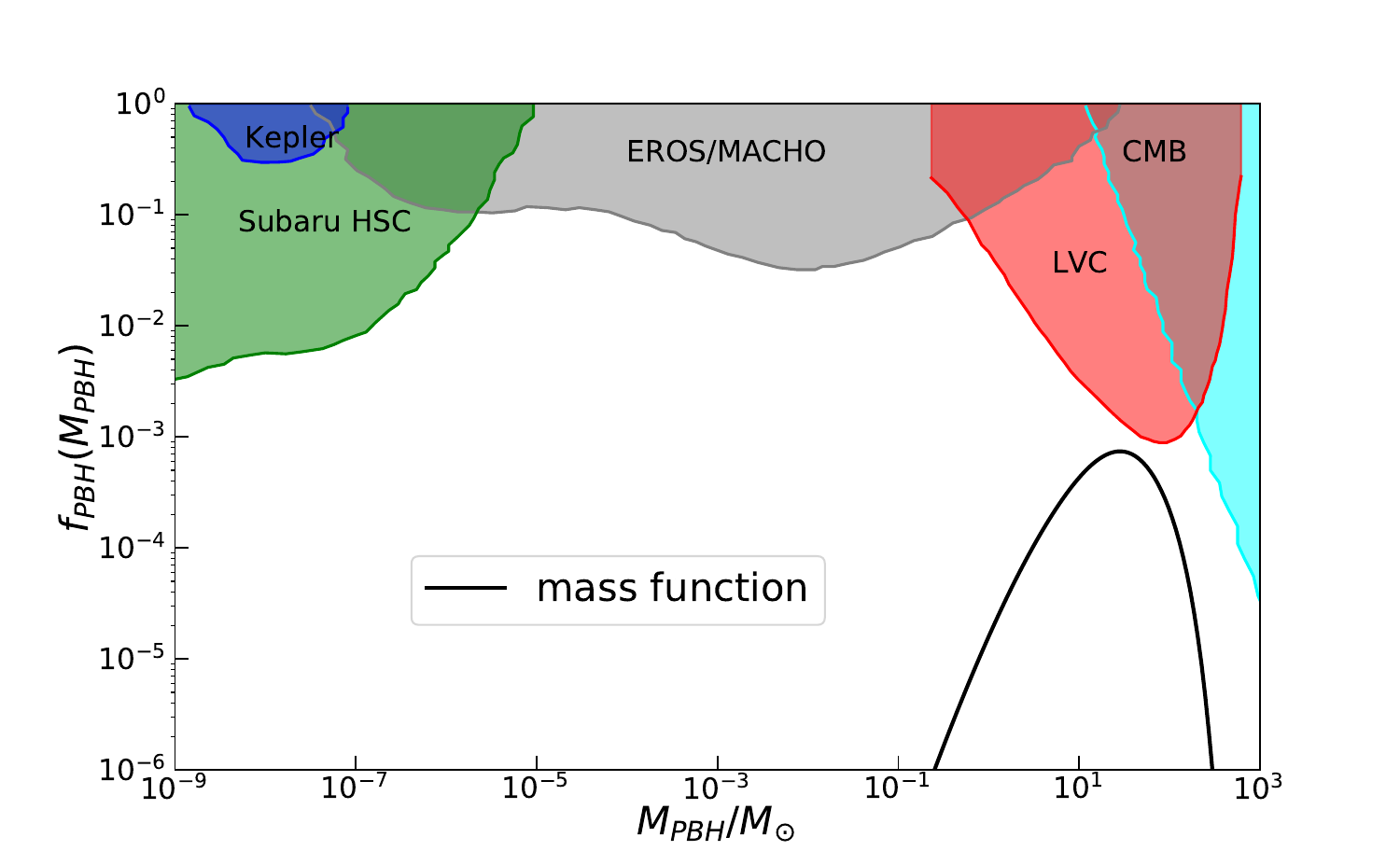}
		\caption{The corresponding  PBHs mass function.
			The shaded regions show the observational constraints on the PBH abundance:
			the cyan region from accretion constraints by CMB \cite{Ali-Haimoud:2016mbv,Poulin:2017bwe},
			the red region from LIGO-Virgo Collaboration measurements \cite{Ali-Haimoud:2017rtz,Raidal:2018bbj,Vaskonen:2019jpv,DeLuca:2020qqa,Wong:2020yig,Hutsi:2020sol},
			the gray region from the EROS/MACHO \cite{Tisserand:2006zx},
			the green region from microlensing events with Subaru HSC \cite{Niikura:2017zjd},
			the blue region from the Kepler satellite \cite{Griest:2013esa}.}
		\label{pic:pbh}
	\end{figure}

	By using the numerical results of the power spectrum of the primordial curvature perturbations,
	we obtain the PBHs mass function shown in Fig. \ref{pic:pbh}.
	The PBHs mass and the mass function at the peak  are
	\begin{equation}\label{pbh:mass12}
		M_\text{PBH}^\text{peak}=29 M_{\odot},\quad f( M_\text{PBH}^\text{peak})=7.40\times 10^{-4},
	\end{equation}
	and the fraction  of the PBHs in dark matter is
	\begin{equation}\label{fpbh:all}
		f_{\text{PBH}}=1.73\times 10^{-3}.
	\end{equation}
	Therefore, for the case  \eqref{para:dq}, we successfully produce the PBHs which can account for the LIGO-Virgo events,
	and the mass function is consistent with all the observational constraints.

\begin{figure}[htp]
		\centering
		\includegraphics[width=0.45\columnwidth]{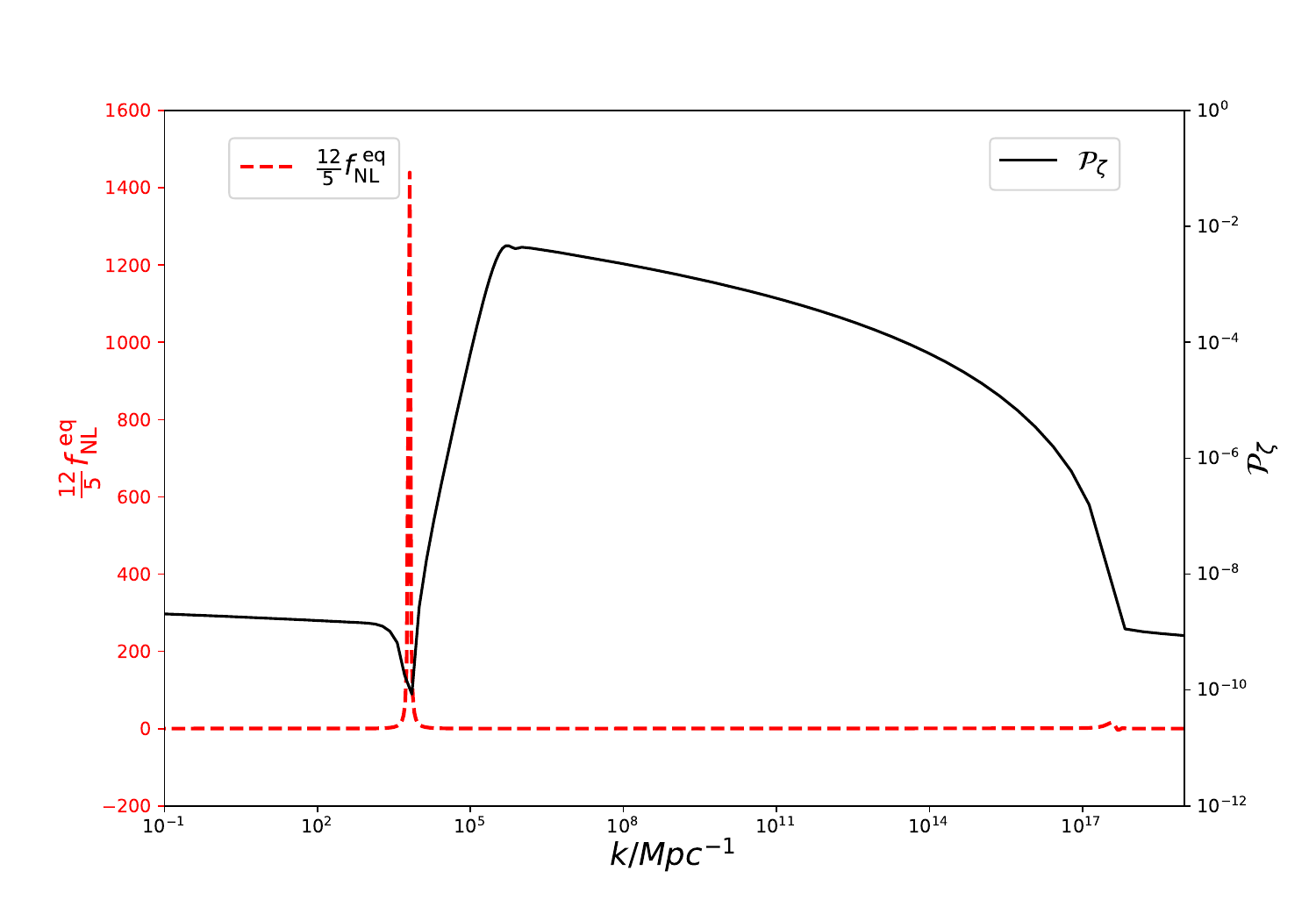}
		\includegraphics[width=0.45\columnwidth]{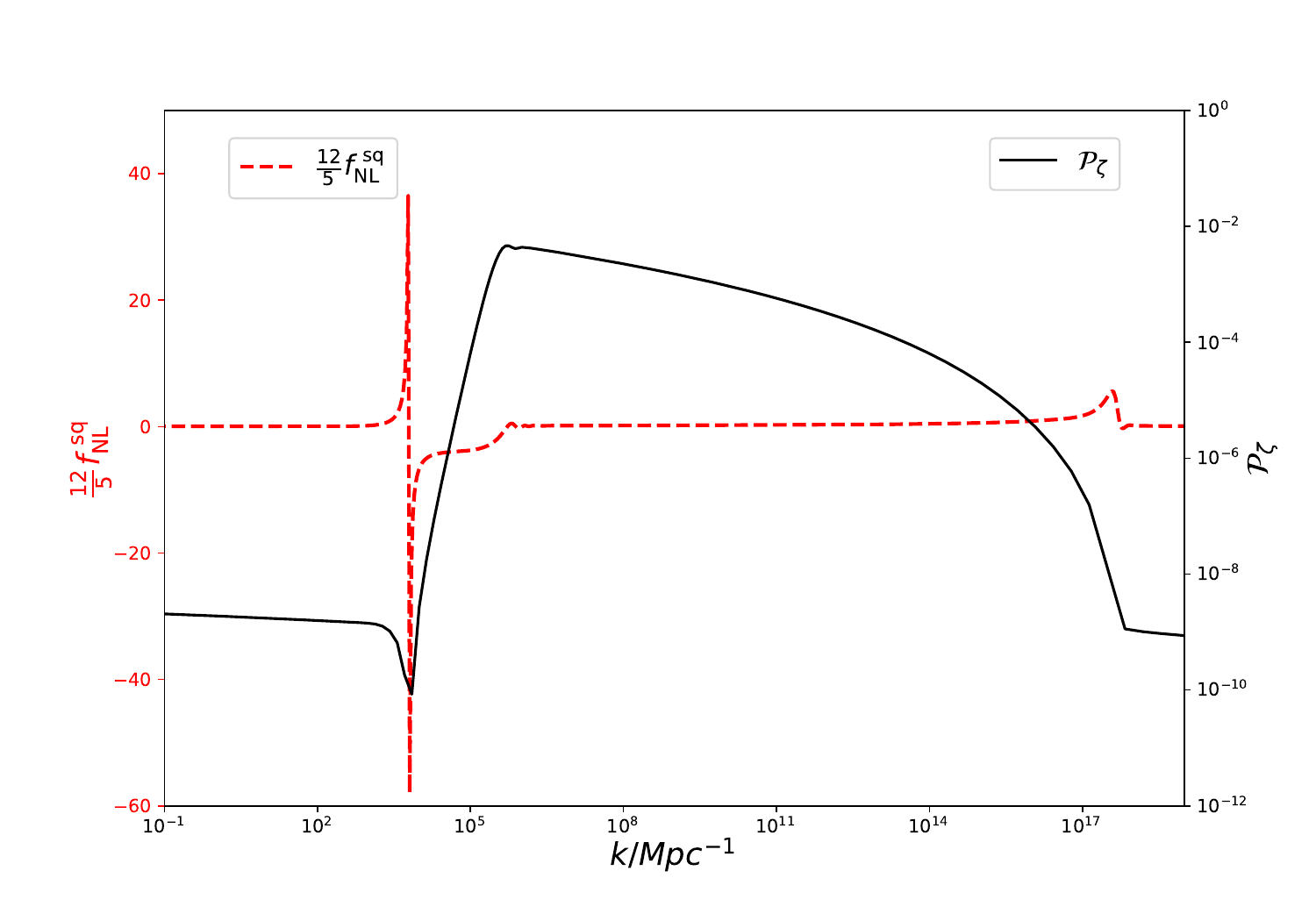}
		\caption{The numerical results for  the non-Gaussianity parameter $f_\text{NL}$ (red dashed line) along with the  scalar power spectrum (black solid line). The left panel shows the results in the equilateral limit  and the right panel shows the results in the  squeezed limit. }
		\label{fig:fng}
	\end{figure}

	There may be some concern that the large enhancement on the power spectrum causes
	large non-Gaussianities which affect the formation of PBH DM.
	The non-Gaussianity parameter $f_\text{NL}$ is 
	\begin{equation}
		f_\text{NL}(k_1,k_2,k_3)=\frac{5}{6}\frac{B_\zeta(k_1, k_2,k_3)}
		{P_\zeta(k_1)P_\zeta(k_2)+P_\zeta(k_1)P_\zeta(k_3)+P_\zeta(k_2)P_\zeta(k_3)},
	\end{equation}
	where $P_\zeta(k)=2\pi^2 \mathcal{P}_\zeta/k^3$ and the definition of the  bispectrum $B_\zeta$ is
	\begin{equation}
		\langle    \hat{\zeta}_{k1} \hat{\zeta}_{k_2} \hat{\zeta}_{k_3} \rangle=(2\pi)^3\delta^3(\bm{k}_1+\bm{k}_2+\bm{k}_3) B_\zeta(k_1,k_2,k_3),
	\end{equation}
	with  $\hat{\zeta}$ being the quantum operator of the curvature perturbation $\zeta_k$.
	The influence of the non-Gaussianities on the abundance of  PBHs can be characterized by the parameter
	\begin{equation}
		\mathcal{J}=\frac{3}{20\pi}f_\text{NL}(k_\text{peak},k_\text{peak},k_\text{peak})
		\sqrt{\mathcal{P}_\zeta(k_\text{peak})}.
	\end{equation}
	If $\mathcal{J}\ll1$,  the  effect of   non-Gaussianities of the curvature perturbation on PBH abundance can be negligible.
	The numerical results for the non-Gaussianity parameter $f_\text{NL}$ are shown in Figure \ref{fig:fng}.  
	At the peak of the power spectrum, the non-Gaussianity parameter $f_\text{NL}$ is small, so the effect of non-Gaussianities in our model is negligible.
	
\subsection{The  NANOGrav constraints and SIGWs} 
\begin{figure}[htp]
		\centering
	 \includegraphics[width=0.45\columnwidth]{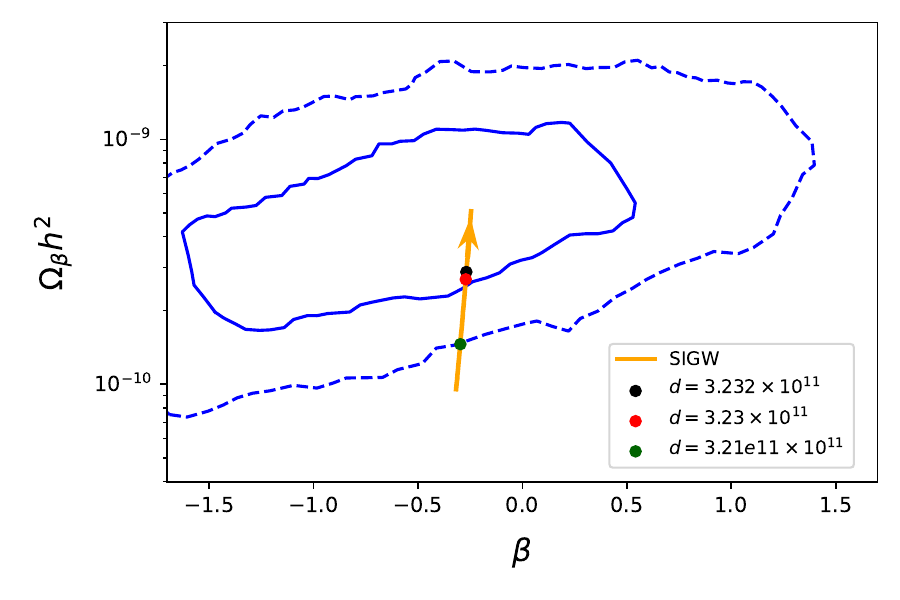}
 	\includegraphics[width=0.45\columnwidth]{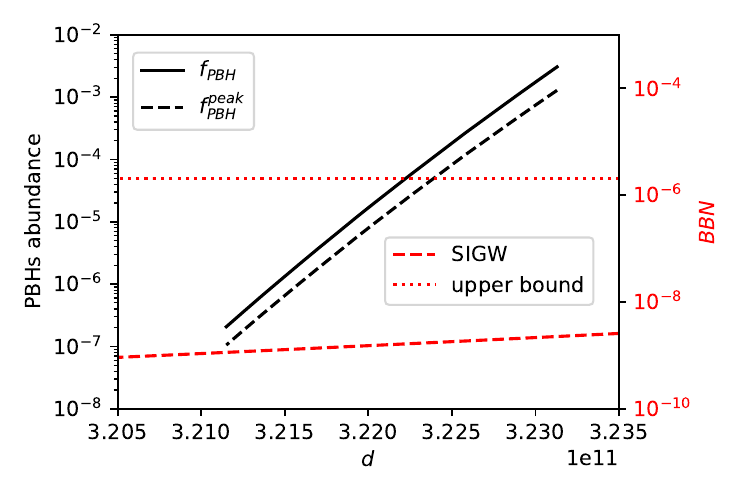}
		\caption{Left panel:  $1\sigma$ (solid blue) and $2\sigma$ (dashed blue) constraints on $\Omega_\beta h^2$  ($f_0=5.5$nHz) and power index $\beta$ indicated by the NANOGrav results with the $5$-frequency power law model \cite{NANOGrav:2020bcs}.  The orange line are the results from our model with different $d$ which increases along the arrow direction. Right panel:  the black  solid line denotes the   fraction  of the PBHs in dark matter, the black dashed line denotes  the PBHs mass function at the peak, the red dashed  line denotes the BBN contributions from the SIGWs, the red dotted line denotes the upper limit of the BBN constraints \eqref{BBN:constraints}. }
		\label{pic:betao}
	\end{figure}

In this section, we apply the NANOGrav experimental results to  obtain the constraints on the only undetermined parameter $d$ in the peak function \eqref{gfuncng}.
The results from the NANOGrav 12.5yrs  data \cite{NANOGrav:2020bcs} are expressed in terms of the characteristic strain  $h_c(f)$  in the narrow frequency range  $f/ \text{Hz} \in (2.5\times 10^{-9},1.2\times 10^{-8})$ with the power-law form 
\begin{equation}\label{pl:strain}
	h_c=A_\alpha\left(\frac{f}{f_\text{yr}}\right)^\alpha,
\end{equation}	
which can be related to the energy density of the GWs by
\begin{equation}\label{gw:strain}
	\Omega_\text{GW}(f)=\frac{2\pi^2}{3H_0^2}f^2h_c^2(f).
\end{equation}
By combining equations \eqref{pl:strain} and  \eqref{gw:strain}, the  predicted energy density of SIGWs in our model can be expanded as
\begin{equation}
	\Omega_\text{GW}(f)=\Omega_\beta\left(\frac{f}{f_0}\right)^{\beta},
\end{equation}
with $f_0=5.5$ nHz \cite{Vaskonen:2020lbd} and
\begin{equation}\label{omega:beta}
	\Omega_\beta=\frac{2\pi^2}{3H^2_0}f_\text{yr}^2A_\alpha^2\left(\frac{f_0}{f_\text{yr}}\right)^\beta,\quad \beta=2\alpha+2.
\end{equation}
Using  equation \eqref{omega:beta}, we can transfer  the experimental results of $A_\alpha$ and $\gamma=3-2\alpha$ to the constraints on the amplitude of SIGWs $\Omega_\beta h^2$ and the power index $\beta$, which  are shown in the left panel of Figure \ref{pic:betao}.  
The solid and dashed blue contours show the $1 \sigma$ and $2\sigma$ constraints on  the  amplitude $\Omega_\beta h^2$ ($f_0=5.5$nHz) \cite{Vaskonen:2020lbd} and power index $\beta$ indicated by the NANOGrav results \cite{NANOGrav:2020bcs}, respectively. The  orange line is the results of the SIGWs  amplitude $\Omega_\beta h^2$ and power index $\beta$ from our model where the parameter $d$  increases along  the arrow direction. The lower limit  of $d$ satisfying the NANOGrav results is $d=3.21\times 10^{11}$ and  marked by the green dot; the  case  \eqref{para:dq}  is  marked by the red dot; the upper limit, from the constraints on the PBHs mass function, is $d=3.232\times10^{11}$ and marked by the black dot. Therefore, the NANOGrav signal and the constraints on the PBHs require  
\begin{equation}\label{para:drange}
    3.21\times10^{11}\leq d\leq3.232\times 10^{11}.
\end{equation}
The black dashed  line in the  right panel of Figure \ref{pic:betao}  denotes the PBHs mass function at the peak and satisfies $1.05\times 10^{-7}\leq f_\text{PBH}^\text{peak}\leq1.29\times 10^{-3}$, the  black  solid line denotes the   fraction  of the PBHs in dark matter and satisfies $2.07\times 10^{-7}\leq f_\text{PBH}\leq3.05\times 10^{-3}$. For the allowed parameter region \eqref{para:drange}, the masses of the PBHs  at the peak  are around $30M_\odot$,  the scalar tilt and tensor-to-scalar ratio are the same with equation \eqref{cmb:nsr}, and the $e$-folding numbers satisfy $59\leq N\leq 65$.

During BBN epoch, the SIGWs contribute to the total energy of extra relativistic species,  and the energy  density of the SIGWs should satisfy the following constraints \cite{Kuroyanagi:2020sfw} 
\begin{equation}\label{BBN:constraints}
	\int _{f_1}^{f_2} d(\ln f)\Omega_\text{GW} (f)h^2\leq 5.6\times 10^{-6} \left(N_\text{eff}^\text{(upper)}-3.046\right),
\end{equation}
where $N_\text{eff}^\text{(upper)}=3.41$ \cite{Cyburt:2015mya} is the upper bound on the effective number of relativistic degrees of  freedom. The lower limit of the integral $f_1=10^{-10}$ Hz \cite{Kuroyanagi:2020sfw} is the frequency of the mode entering the horizon at the BBN epoch, the upper limit of the integral $f_2=10^{7}$Hz  \cite{Kuroyanagi:2014nba} is determined  by the Hubble parameter at the end of inflation. 
The contributions to  the total energy of extra relativistic species from the  SIGWs of our model are shown in 
the right panel of the Figure \ref{pic:betao} and denoted by the red dashed line which  is consistent with the BBN constraints \eqref{BBN:constraints}  denoted by the red dotted   line.

\begin{figure}[htp]
		\centering
		\includegraphics[width=0.65\columnwidth]{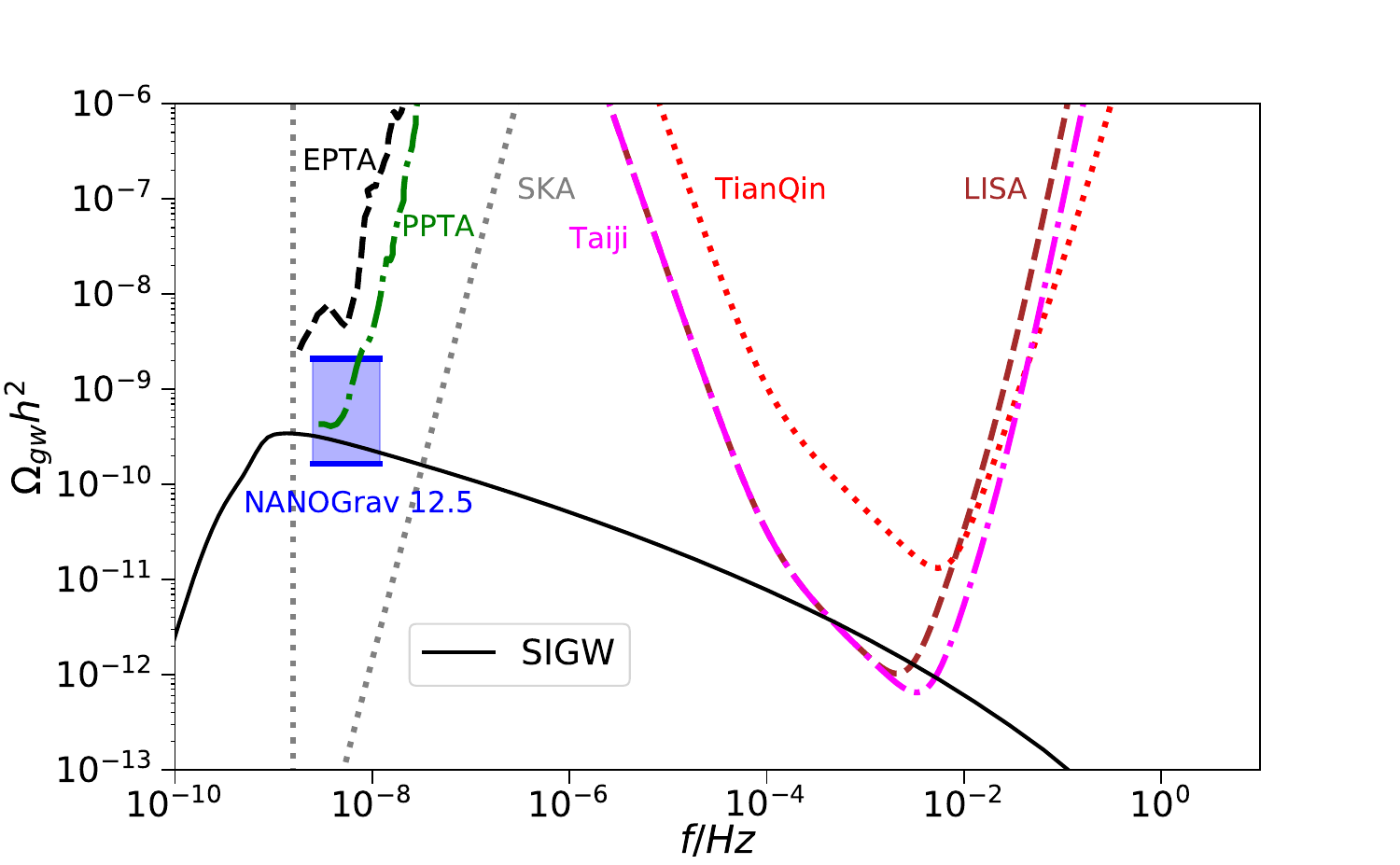}
		\caption{The scalar induced secondary GWs from the Higgs field for the case \eqref{para:dq}.
			The black dashed curve denotes the EPTA limit \cite{Ferdman:2010xq,Hobbs:2009yy,McLaughlin:2013ira,Hobbs:2013aka,Lentati:2015qwp},
			the green dot-dashed curve denotes the PPTA limit \cite{Shannon:2015ect},
			the gray dotted curve denotes the SKA limit \cite{Moore:2014lga},
			the red  dotted curve in the middle denotes the TianQin limit \cite{Luo:2015ght},
			the magenta dot-dashed curve shows the Taiji limit \cite{Hu:2017mde},
			the brown dashed curve shows the LISA limit \cite{Audley:2017drz}.}
		\label{pic:gw}
	\end{figure}

For the case  \eqref{para:dq}, we display the energy density  of the SIGWs as a function of the frequency  in Figure \ref{pic:gw}. The frequencies of the SIGWs cover from nanohertz to millihertz.
Around the nanohertz, the energy density  of the SIGWs is consistent
with the $2 \sigma $ region of the NANOGrav signal; 
around the millihertz, the SIGWs can be detected
by space-based GW detectors such as  Taiji and LISA.
This means that the NANOGrav signal may originate from the Higgs field and may also 
be detected by space-based GW detectors in the future.
	
\section{Conclusion}
The NANOGrav signal can be explained by the  gravitational waves induced from large scalar
perturbations at small scales during the radiation dominated epoch.
By introducing a noncanonical kinetic term with a high peak in  inflation models, the amplitude of the power spectrum of the
primordial curvature perturbations can be enhanced to generate SIGWs and  produce PBHs  at small scales while keeping small to satisfy the Planck 2018 observational data at large scales. 
With this mechanism,  the Higgs field successfully produces PBHs accounting for the  LIGO-Virgo  events and generates the SIGWs explaining  the  NANOGrav signal. 
	
The PBHs observational constraints  and 	the  NANOGrav 12.5yrs experimental results require the parameter in the peak function $G_p(\phi)$  to satisfy  $3.21\times  10^{11}\leq d \leq 3.232\times 10^{11}$. Within this parameter range, the masses of the PBHs at the peak are around   $M_\text{PBH}^\text{peak}\approx30  M_{\odot}$, the total fraction of the PBHs in dark matter satisfies $2.07\times 10^{-7}\leq f_\text{PBH}\leq3.05\times 10^{-3}$.  The scalar tilt and tensor-to-scalar ratio among these models are almost the same, which is  $n_s= 0.966$ and $r=0.039$, and consistent with Planck 2018 observational data,  the $e$-folds for the pivot scale $k_*=0.05\text{Mpc}^{-1}$ satisfy  $59\leq N\leq65$.
The energy densities of the corresponding SIGWs are consistent with  the $2 \sigma $ region of the NANOGrav signal, and these SIGWs    can  be also detected by the  future space-based GW detectors such as Taiji and LISA.

In conclusion, the NANOGrav signal  and the BHs in LIGO-Virgo events  can both originate from the Higgs field, and the NANOGrav signal may also  be detected by the space-based GW detectors in the future.

\acknowledgments
This work is supported by the National Natural Science Foundation of China under Grants
Nos. 11633001, 11920101003 and 12021003, the Strategic Priority Research Program of the Chinese Academy of Sciences,
Grant No. XDB23000000 and the Interdiscipline Research Funds of Beijing Normal University.
Z. Y. thanks Fengge Zhang for the useful discussion.
	


\providecommand{\href}[2]{#2}\begingroup\raggedright\endgroup

\end{document}